# Does the World Bank's Ease of Doing Business Index Matter for FDI? Findings from Africa

Bhaso Ndzendze[1]

Department of Politics and International Relations, University of Johannesburg

**Abstract**

This paper investigates whether foreign investment (FDI) into Africa is at least partially responsive to World Bank-measured market friendliness. Specifically, I conducted analyses of four countries between 2009 and 2017, using cases that represent two of the highest scorers on the bank's *Doing Business* index as of 2008 (Mauritius and South Africa) and the two lowest scorers (DRC and CAR), and subsequently traced all four for growths or declines in FDI in relation to their scores in the index. The findings show that there is a moderate association between decreased costs of starting a business and growth of FDI. Mauritius, South Africa and the DRC reduced their total cost of starting a business by 71.7%, 143.7% and 122.9% for the entire period, and saw inward FDI increases of 167.6%, 79.8% and 152.21%, respectively. The CAR increased the cost of starting businesses but still saw increases in FDI. However, the country also saw the least amount of growth in FDI at only 13.3%.

**Introduction**

This paper asks whether the *Ease of Doing Business* index is a predictor of movements in foreign direct investment (FDI) influx into African states. Particularly, the paper seeks to infer whether there is a causal relationship between the various countries' scores on the cost of doing business variable of the *Doing Business* index and subsequent growths or declines in FDI to any given country. Essentially, the paper adopts as an independent variable the changes in scores in the World Bank's Doing Business index over 2009 to 2017. This score is represented as the cost of starting a business as a percentage per capita, and thus could range from 0 to well over 100%. Importantly, this score may or may not change in each consecutive year. On the other hand, the dependent variable for this paper is the subsequent growth or decline in the influx of FDI on an annual basis. Specifically, this is operationalized as an increase or decrease in inbound foreign direct investment stocks in monetary terms (in US dollars). The countries being studied here are Mauritius, South Africa, the Central African Republic (CAR) and the Democratic Republic of the Congo (DRC). The conditions for case selection which resulted in the inclusion of these four countries necessitated a comparative analysis of the two highest and two of the lowest ranked African states in the index as of 2008 – at number 27 and 35 in the ranking, Mauritius and South Africa came first and second among African countries, respectively, while at number 177 and 178, CAR and the DRC were ranked second-last and last among African countries and in the global ranking overall, respectively.

---

[1] Associate Professor: International Relations. E-mail: bndzendze@uj.ac.za

The overall findings from the research display a strong association between decreased costs of starting a business and FDI influx. But rather than it being the case that the high rankers by virtue of this status saw the most influx of FDI, it was rather the case that incidences of reform (i.e., decreases in their cost of starting a business) by them and the DRC gained the most increases in FDI. Mauritius reduced its total cost of starting a business by 71.7% for the entire period, and saw a 167.6% in total FDI. South Africa reduced its cost of starting a business by 143.7% and saw a 79.8% increase in total FDI. And the DRC cut its cost of starting a business by 122.9% and saw a 152.21% increase in FDI.

However, there were some mixed results, indicating that the movements were not universally correlated and linear. The case of CAR, which saw increased in inward FDI for all the years but only reformed for five incidences out of nine, exhibits this trait. Indeed, the country increased its cost of starting a business by 49%, and still saw an increase in FDI. However, the country saw the least growth in FDI, growing only by 13.3% and in raw figures, the country obtained the least FDI for the entire period; growing to only US$651-million by 2017, from US$407-million in 2008. Thus, among the low rankers, in the entire timeframe the DRC proved the most active reformer and CAR overtook it as the *lower* ranking African country in the rankings. As the second-highest cutter in cost of doing business (after South Africa), the country saw the second-highest growth in FDI (after Mauritius) in the entire dataset.

The first section of the paper will conduct a literature review, contextualising and describing the *Doing Business* index, as well as conceptualising foreign direct investment. The section also links the *DB* index with the concept of FDI. The paper then details the methodology to be applied in the case studies. In the fourth section, the paper cascades the results from the data, and subsequently analyses the findings on a country-by-country basis in the fifth section. These country-specific findings are discussed in tandem with one another in the sixth section of the paper. Finally, the paper concludes with an assessment of the emergent research questions for future research.

**Literature Review**

***What is the Ease of Doing Business Index?***

The World Bank has set itself apart by not only functioning as a global financial institution, but also as an idea generator, particularly on the issue of economic development and its antecedents. To these ends, the bank has produced a library of publications in the form of working papers, books and reports which communicate these ideas. Among these has been the World Bank's *Doing Business* report, published annually since the year 2003. The principal idea behind the report and its methodology, as will be seen, is the necessity of openness, minimal state intrusion in enterprises and private sector-led development. Notably, the report is one among many similar such reports and indices, including the younger World Economic Forum's (WEF) Competitiveness Report published since 2008.

Published annually, between 2003 and 2013 the report ranked countries from number 1 to number 185, and since 2014, with the inclusion of Libya, Myanmar, San Marino and South Sudan, has ranked a total of 189 countries. The report has been embraced by a litany of important actors in the global community, including especially the governments of developing countries eager to retain as well as attract new foreign investment. In India, it was a noteworthy development in 2018 when the country jumped 23 places and come to rank at 77[th] place.[1] In China, the same kind of news was welcomed in the same year when the country jumped to the 46[th] place.[2] In much of the developing the world, including in Africa, the orthodoxy of the World Bank's *Doing Business* report, despite the continent's history with its ideas including, for example, the structural adjustment programs, which are blamed for underdevelopment in many quarters, have found acceptance.[3] The World Bank has worked closely with numerous willing governments across all developing regions (East and South Asia, the Middle East and North Africa, Eastern Europe and Central Asia, Sub-Saharan Africa and Latin America) in bringing them closer to perfect scores on the ranking.

> To ensure the coordination of efforts across agencies, such economies as Brunei, Darussalam, Colombia and Rwanda have formed regulatory reform committees, reporting directly to the president. These committees use the *Doing Business* indicators as one input to inform their programs for improving the business environment. More than 45 other economies have formed such committees at the interministerial level. Since 2003 governments have reported more than 530 regulatory reforms that have been informed by *Doing Business*.[4]

But what sort of markers does the *Doing Business* seek to identify and champion? Principally, *Doing Business* presents 'quantitative indicators' on business regulations and the protection of property rights. These are 10 – ostensibly semi-comprehensive – regulations which arguably affect '10 stages of a business's life.'[5] These are:

- starting a business;

- dealing with licenses;

- employing workers;

- registering property;

- getting credit;

- protecting investors;

- paying taxes;

- trading across borders;

- enforcing contracts; and

- closing a business.

There are numerous other, uneven, sub-categories within this broad 10. For example, in the first regulation (starting a business), there are the following indicators which are measured and scored accordingly:

(1) The number of procedures required to start a business, which is measured in a raw number;

(2) Time normally taken to register a new business, which is measured in the number days;

(3) The cost of starting a business, which is calculated as a percentage (%) of income per capita;

(4) Minimum capital required, which is calculated as a percentage of income per capita.

(Incidentally, this paper will be interested in measuring the cost of starting a business as an independent variable in bringing about increased FDI to the countries under analysis.) The data for each year are released on a retrospective basis; for example, 'data in *Doing Business 2008* are current as of June 1, 2007.'[6] According to the World Bank itself, 'the indicators are used to analyze economic outcomes and identify what reforms have worked, where, and why.'[7] This interest in reform stemming from the index is reflected in the research papers which tend to accompany each annual report itself. By the report's own analysis, the tide has indeed been towards more openness. As the 2014 report noted,

> in 2005 the time to start a business in the economies ranking in the worst quartile on this indicator averaged 113 days. Among the best 3 quartiles it averaged 29 days. Today that gap is substantially narrower. While the difference is still substantial at 33 days, it is considerably smaller than the 85 days in 2005.[8]

'Reformers' are those countries which would have on the previous year 'simplified business regulations, strengthened property rights, eased tax burdens, increased access to credit and reduced the cost of exporting and importing.'[9] In the 2008 report, for example, 'across regions, Eastern Europe and Central Asia reformed the most, followed by South Asia and rich countries.'[10] All the while, 'Latin America reformed the least.'[11] Accordingly, then, the World Bank argues that the *DB* index is a reliable predictor of heightened commercial activity. As it noted in relation to one country,

> Two years ago registering a property in Croatia took 956 days. Now it takes 174. Croatia also sped company start-up, consolidating procedures at the one-stop shop and allowing pension and health services registration online. Two procedures and 5 days were cut from the process. Credit became easier to access: a new credit bureau got off the ground, and a unified registry now records all charges against movable property in one place.[12]

The World Bank saw this as having resulted in improved investment levels in the country: 'in the first 2 months €1.4 billion of credit was registered. Finally, amendments to the Croatian insolvency law introduced professional requirements for bankruptcy trustees and shortened timelines.'[13] Stellar performers on the continent have included Mauritius, which regularly tops the list, along with Rwanda, Zambia, and Egypt. In 2008,

> Egypt's reforms went deep. They made starting a business easier, slashing the minimum capital requirement from 50,000 Egyptian pounds to 1,000 and halving start-up time and cost. Fees for registering property were reduced from 3% of the property value to a low fixed fee. With more properties registered and less evasion, revenue from title registrations jumped by 39% in the 6 months after the reform. New one-stop shops were launched for traders at the ports, cutting the time to import by 7 days and the time to export by 5.[14]

In many African countries, while there is a burgeoning entrepreneurial trend, especially among youths,[15] who also make up the majority of the continent's population,[16] there is also a clear absence of large middle and upper middle class to champion investment.[17] This is crucial and comes to mean that discussions of attracting investment nominally mean attracting foreign investors to these economies, many of whom are still reliant on foreign funding in other forms, particularly aid and remittances.[18] Thus, in the absence of domestic capital, the focus is on foreign investment, traditionally western, but increasingly dominated by Chinese financiers. Thus, this paper is interested in the utility of the *Doing Business* index as a predictor of increases in foreign investment in African countries. Having defined the first portion of that question, it is conceptually necessary to disambiguate the latter portion; namely, what is foreign direct investment?

### *China and foreign direct investment*

Like many concepts in International Relations, FDI can be a difficult concept to define in axiomatic terms. Further, interpretations of the implications of FDI have evolved over time: 'while they were perceived primarily as a threat to national economic development from the 1950s to the 1980s, they came to be gradually re-interpreted as a sign of economic success in the 1990s.'[19] However, some core descriptive characteristics simplify such a task. Essentially, FDI is defined both as process and its end-goal. Specifically, it is the act of disbursing funds by an individual or corporation in another country for the purposes of obtaining profits after a given time frame. As the United Nations Conference on Trade and Development (UNCTAD) more surgically puts it:

> FDI is defined as an investment involving a long-term relationship and reflecting a lasting interest and control by a resident entity in one economy (foreign direct investor or parent enterprise) in an enterprise resident in an economy other than that of the foreign direct investor (FDI enterprise or affiliate enterprise or foreign affiliate).[20]

This naturally involves some vulnerability and exposure for the investor. There exists the risk, for example, of the seizing of multinational corporations' properties by the domicile country's government, or at least the lack of capacity of the government to prevent such a seizing by non-state actors within the country.[21] There also exists the risk of arbitrary laws being applied to some properties and investments and not others. Further, the associated costs of opening and operating a new enterprise may prove profit-reducing for some would-be investors after the fact. Further, the OLI framework identifies the necessary factors which may determine whether a multinational corporation pursues a foreign investment. 'OLI,' which denotes 'Ownership, Location, and Internalization,' identifies these as the three potential sources of advantage that may determine whether a firm decides to become a multinational. 'Ownership advantages address the question of why some firms but not others go abroad, and suggest that a successful MNE [multinational enterprise] has some firm-specific advantages which allow it to overcome the costs of operating in a foreign country.'[22] Secondly, location refers to 'the question of *where* an MNE chooses to locate.'[23] The third and final aspect includes internalization advantages influence how a firm chooses to operate in a foreign country, trading off the savings in transactions, holdup and monitoring costs of a wholly-owned subsidiary, against the advantages of other entry modes such as exports, licensing, or joint venture.[24]

The key aspects of this theory are the fact that it focuses on the incentives facing individual firms. Nominally, 'the majority of work on the motives of FDI refers to the four main determinants: firstly, the search for natural resources (resource-seeking); secondly, the search for the markets (market-seeking); thirdly, the search for strategic assets (strategic asset-seeking) to acquire technology, managerial capacities, brands, distribution networks and other assets and lastly, the search for efficiency (efficiency-seeking) to exploit economies of scales, or by securing access to cheaper inputs.'[25] Importantly, the very notion of FDI and much of what has been written about it takes it as a given that the funds to be invested traditionally come from certain parts of the world; those with excess capital. These capital-intensive economies are nominally those of the West, especially the US, the UK, Germany and France.[26] However, as of the two decades between 2000 and 2019, China and India have also emerged as active foreign investors;[27] true to form, however, this is also a marker of their having entered the class of emerging power status – and access to some excess capital.[28] As emphasised in the 2009 UNCTAD World Investment Report, companies and funds from a number of Asian economies that are not, or are less, affected by the financial turmoil may maintain an aggressive strategy for overseas investments and become more important actors on the global FDI scene. Furthermore, for many Chinese and Indian companies in particular, the desire to acquire undervalued assets (such as mineral deposits, technologies, brand names and/or distribution networks) during the global and financial crisis may boost Asian Investments in developed countries.[29]

Within Africa, also, South Africa is a long-standing investor (with its capital mostly present in Nigeria, Lesotho, Botswana, Angola, and Rwanda), though one which also remains a recipient and an active seeker of FDI.[30]

*The* DB *index and FDI*

To make its data comparable to as many countries as possible, *DB* indicators 'refer to a specific type of business−generally a limited liability company operating in the largest business city.'[31] (World Bank, 2008: 3). In line with the *DB* index's conceptualization, it is necessary for any would-be investor to obtain a certain level of ease as well as security of the intended country of investment. Signals are crucial in this regard.[32] 'FDI should be empowered through appropriate economic and political governance by recipient countries.'[33] The *DB* index in this way allows us to measure and quantify any country's standing in this regard.[34] The 10 stages which the index speaks to, as detailed in the previous section, also speak directly to the guarantees needed by the would-be investor. In a post-2008/09 financial crisis world, these investors have been relatively reluctant to invest in potentially investor unfriendly markets.[35] Many – though, crucially, not all – African states are at the extreme ends of this investor friendliness spectrum; with some being perceived as major risks and others as newly burgeoning markets with *growth* potential than their western counterparts, many of whom are facing slowdowns or stagnations.[36]

In line with this literature, this paper seeks to ask the following question: are international investments in these four African countries at least partially driven by market friendliness of countries on the continent? The methodology for the analysis being conducted is detailed in the following section.

**Methodology**

*Variables and case selection*

The paper seeks to infer the causal relationship between movements in various countries' scores on the cost of doing business variable of the *Doing Business* index and subsequent growths or declines in inward FDI over the period from 2009 to 2017. This score serves as the independent variable in this paper is represented as the cost of starting a business as a percentage per capita, and thus could range from 0 to over 100%. Importantly, this score may or may not change in each consecutive year. On the other hand, the dependent variable for this paper is the subsequent growth or decline in the influx of FDI on an annual basis. Specifically, this is operationalized as an increase or decrease in inbound foreign direct investment stocks in monetary terms (i.e., in US dollars). The countries being studied here are Mauritius, South Africa, the Central African Republic and the Democratic Republic of the Congo. The conditions for case selection which resulted in the inclusion of these four countries to represent the two highest ranking African states in the index and two of the lowest ranked African states as of 2008.

*Method of inquiry*

Since the paper will be primarily interested in detecting whether those countries which went on to reduce their costs of doing business scores also saw increases in FDI, our method of inquiry is a before-and-after analysis across a time series, and features both a variant independent variable, as some states may not affect changes onto their scores, as well as variance on the dependent variable as the paper is open to there being no changes in FDI influx since nothing guarantees this and as the correlation may not necessarily occur in uniform levels across all the countries.[37]

*Data*

The data on the countries' scores on the *Doing Business* index and subsequent costs of starting a business in the countries was obtained from the World Bank's reports between 2008 and 2017. The data on the total FDI influx into the four countries was obtained from the United Nations Conference on Trade and Development's Data Center.

**Results**

The four tables below represent the findings made from the data in raw and statistical terms. Four each country, the first column represents the successive years of study from 2009 through 2017. The second and third columns capture the FDI data for the respective years and the subsequent growth or decline from the previous year, respectively. The fourth and fifth columns capture the data for the countries' scores on the index per year and the movement in that score in the successive years, respectively.

**Table 1: Mauritius' comparative FDI stock and *DB* scores, 2009-2017**

| Year | Total FDI stock (in US$ billions) | Comparative FDI movement (in %) | Index score | Movement in cost of starting a business (in %) |
|------|------|------|------|------|
| 2009 | 3.016 | 84.8 | 5.0 | -5.6 |
| 2010 | 4.658 | 54.4 | 4.1 | -18 |
| 2011 | 2.999 | -35.6 | 3.8 | -7.3 |
| 2012 | 3.218 | 7.3 | 3.6 | -5.2 |
| 2013 | 4.345 | 35.0 | 3.3 | -8.3 |
| 2014 | 3.497 | -19.5 | 3.6 | 9.0 |

| | | | |
|------|-------|------|-------|
| 2015 | 4.275 | 22.2 | 2.1 | -41.6 |
| 2016 | 4.499 | 5.2  | 2.0 | -4.7 |
| 2017 | 5.122 | 13.8 | 1.8 | 10 |

Sources: World Bank, *Doing Business 2008*. Washington, D.C.: World Bank and the International Finance Corporation, 2008, pp. 136 and United Nations Conference on Trade and Development Data Center, 'Foreign direct investment: Inward and outward flows and stock, annual', 2019 <https://unctadstat.unctad.org/wds/TableViewer/tableView.aspx?ReportId=96740>. Calculations by author.

**Table 2: South Africa's comparative FDI stock and *DB* scores, 2009-2017**

| Year | Total FDI stock (in US$ billions) | Comparative FDI movement (in %) | Index score | Movement in cost of starting a business (in %) |
|------|-----------|-------|-----|-------|
| 2009 | 138.751 | 65.8  | 6.0 | -15.4 |
| 2010 | 179.565 | 29.4  | 5.9 | -1.6 |
| 2011 | 159.390 | -11.2 | 6.0 | 1.6 |
| 2012 | 163.510 | 2.5   | 0.3 | -95.0 |
| 2013 | 152.123 | -6.9  | 0.3 | 0.0 |
| 2014 | 138.906 | -8.6  | 0.3 | 0.0 |
| 2015 | 126.755 | -8.7  | 0.3 | 0.0 |
| 2016 | 135.454 | 6.8   | 0.3 | 0.0 |
| 2017 | 149.962 | 10.7  | 0.2 | -33.3 |

Sources: World Bank, *Doing Business 2008*. Washington, D.C.: World Bank and the International Finance Corporation, 2008, pp. 150 and United Nations Conference on Trade and Development Data Center, 'Foreign direct investment: Inward and outward flows and stock, annual', 2019 <https://unctadstat.unctad.org/wds/TableViewer/tableView.aspx?ReportId=96740>. Calculations by author.

**Table 3: Central African Republic's comparative FDI stock and *DB* scores, 2009-2017**

| Year | Total FDI stock (in US$ billions) | Comparative FDI movement (in %) | Cost of starting a business (% of income per capita) | Movement in cost of starting a business (in %) |
|------|------|------|------|------|
| 2009 | 0.449 | 10.3 | 232.3 | 13.0 |
| 2010 | 0.511 | 13.8 | 244.9 | 5.4 |
| 2011 | 0.548 | 7.2 | 228.4 | -6.7 |
| 2012 | 0.618 | 12.7 | 175.5 | -23.1 |
| 2013 | 0.619 | 0.1 | 172.6 | -1.6 |
| 2014 | 0.623 | 0.6 | 162.0 | -6.1 |
| 2015 | 0.626 | 0.4 | 226.0 | 39.5 |
| 2016 | 0.633 | 1.1 | 204.0 | -9.7 |
| 2017 | 0.651 | 2.8 | 209.4 | 2.6 |

Sources: World Bank, *Doing Business 2008*. Washington, D.C.: World Bank and the International Finance Corporation, 2008, pp. 114 and United Nations Conference on Trade and Development Data Center, 'Foreign direct investment: Inward and outward flows and stock, annual', 2019

<https://unctadstat.unctad.org/wds/TableViewer/tableView.aspx?ReportId=96740>. Calculations by author.

**Table 4: Democratic Republic of the Congo's comparative FDI stock and *DB* scores, 2009-2017**

| Year | Total FDI stock (in US$ millions) | Comparative FDI movement (in %) | Cost of starting a business (% of income per capita) | Movement in cost of starting a business (in %) |
|------|------|------|------|------|
| 2009 | 6.429 | 11.51 | 435.4 | -10.6 |
| 2010 | 9.368 | 45.7 | 391.0 | -10.1 |
| 2011 | 11.055 | 18.0 | 735.1 | 88.0 |
| 2012 | 14.367 | 29.9 | 551.4 | -24.9 |
| 2013 | 16.465 | 14.60 | 284.7 | -48.3 |
| 2014 | 18.309 | 11.1 | 200.1 | -29.7 |
| 2015 | 19.982 | 9.1 | 30.0 | -85.0 |
| 2016 | 21.187 | 6.0 | 29.3 | -2.3 |
| 2017 | 22.527 | 6.3 | 29.3 | 0.0 |

Sources: World Bank, *Doing Business 2008*. Washington, D.C.: World Bank and the International Finance Corporation, 2008, pp. 112 and United Nations Conference on Trade and Development Data Center, 'Foreign direct investment: Inward and outward flows and stock, annual', 2019 <https://unctadstat.unctad.org/wds/TableViewer/tableView.aspx?ReportId=96740>. Calculations by author.

**Analysis**

This section will test out the hypothesis by assessing the correlations between the two variables. Particularly worth assessing will be the responsiveness of FDI to decreases and increases in the cost of doing business for each country.

*Mauritius*

**Figure 1: Mauritius**

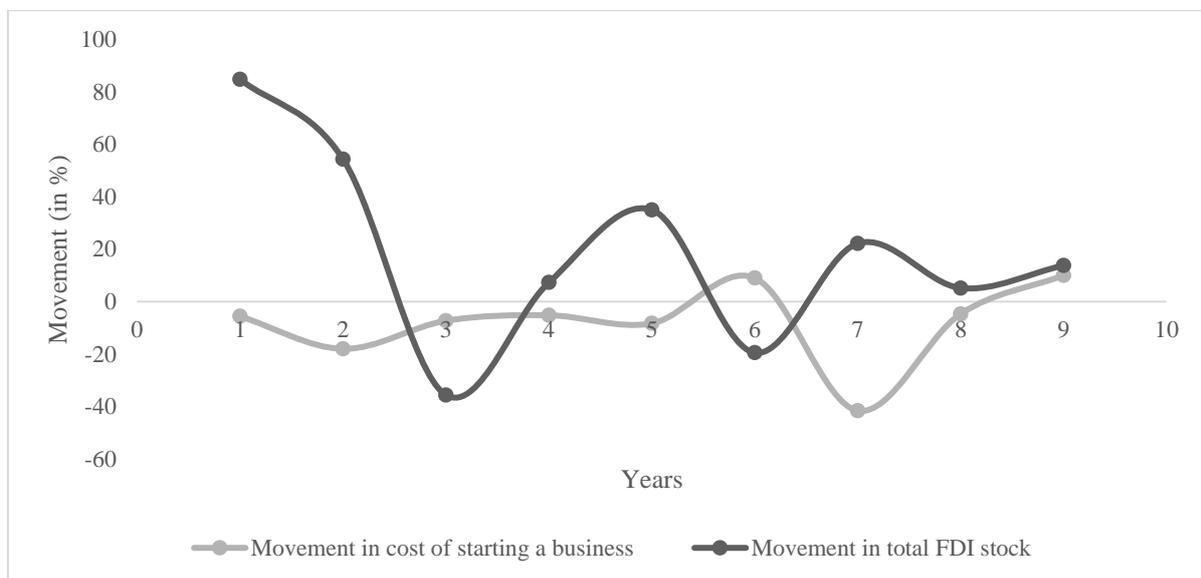

Sources: United Nations Conference on Trade and Development Data Center, 'Foreign direct investment: Inward and outward flows and stock, annual', 2019 <https://unctadstat.unctad.org/wds/TableViewer/tableView.aspx?ReportId=96740> and World Bank, *Doing Business* reports 2008-2017. Washington, D.C.: World Bank and the International Finance Corporation. <http://www.doingbusiness.org/en/reports/global-reports/doing-business-2019>. Calculations by author.

As displayed in Figure 1, Mauritius decreased its cost of doing business for all the years in the dataset with the exception of 2014 and 2017. Throughout the dataset, the variables did observe inverse correlations for seven of the nine years; as the cost of doing business decreased, the total FDI stock increased: in 2009, in 2010, in 2012, in 2013, in 2014, in 2015, and in 2016.

Importantly, while 5 of these six years saw a decrease in cost of starting a business and a growth in FDI, the movement in the opposite direction took place in 2014 wherein the cost of starting a business increased by 9.0%, and there was a decrease in FDI by 19.5%. This is in keeping with the hypothesis, and indicates a causality.

However, there were two years wherein the correlation seemed to not exist. In 2011, there was a lack of inverse correlation, when the cost of starting a business decreased by 35.6%, but there was a *decline* in FDI of 7.3%. In 2017, there was also a lack of inverse correlations in the reverse direction when the cost of starting a business increased by 10% but there was nonetheless an increase in FDI of 13.8%. We can perhaps explain the lack of correlation in 2017 as being due to the fact that even though the increase took place, it was an increase from an already low base of 1.8%.

*South Africa*

**Figure 2: South Africa**

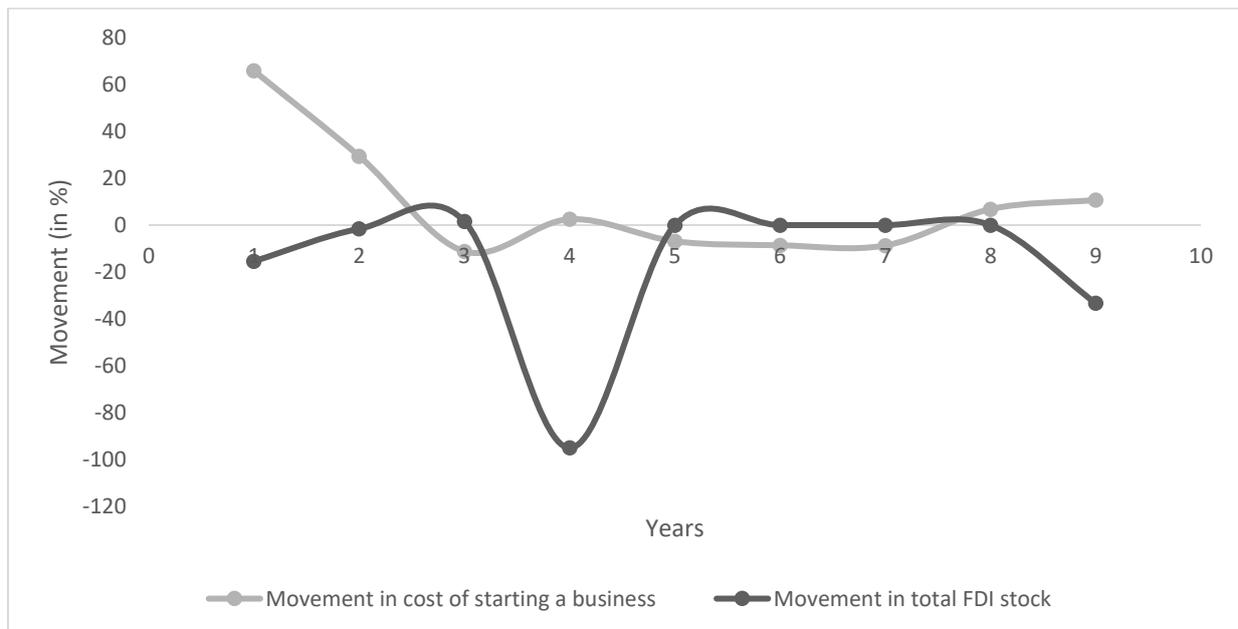

Sources: United Nations Conference on Trade and Development Data Center, 'Foreign direct investment: Inward and outward flows and stock, annual', 2019

<https://unctadstat.unctad.org/wds/TableViewer/tableView.aspx?ReportId=96740> and World Bank, *Doing Business* reports 2008-2017. Washington, D.C.: World Bank and the International Finance Corporation. <http://www.doingbusiness.org/en/reports/global-reports/doing-business-2019>. Calculations by author.

As represented in Figure 2, South Africa decreased its cost of doing business for the years in the dataset with the exception of 2011, and all years in the 2013 to 2016 period. Throughout the dataset, the variables did observe inverse correlations for five of the nine years; as the cost of doing business decreased, the total FDI stock increased: in 2009, in 2010, in 2011, in 2012, and in 2017. Importantly, while 5 of the six years saw a decrease in cost of starting a business and a growth in FDI, the movement in the opposite direction took place in 2011 wherein the cost of starting a business increased by 1.6% and there was a decrease in FDI by 11.2%. This is in keeping with the hypothesis, and indicates a causality.

However, there were two years wherein the correlation seemed to not exist. In the 2013-2016 period, there was a lack of inverse correlation, when the cost of starting a business neither increased nor decreased but there were nonetheless decreases in 2013, 2014 and 2014, and an increase in 2016. This does indicate some spuriousness in the causal account; in other words, that there would be no movement in the cost of starting a business but there would be a movement in the total inward FDI stock to South Africa indicates that the *DB* index, or in the very least this variable, is not sufficient to explain the movement of FDI, at least in regards to South Africa.

*Central African Republic*

**Figure 3: Central African Republic**

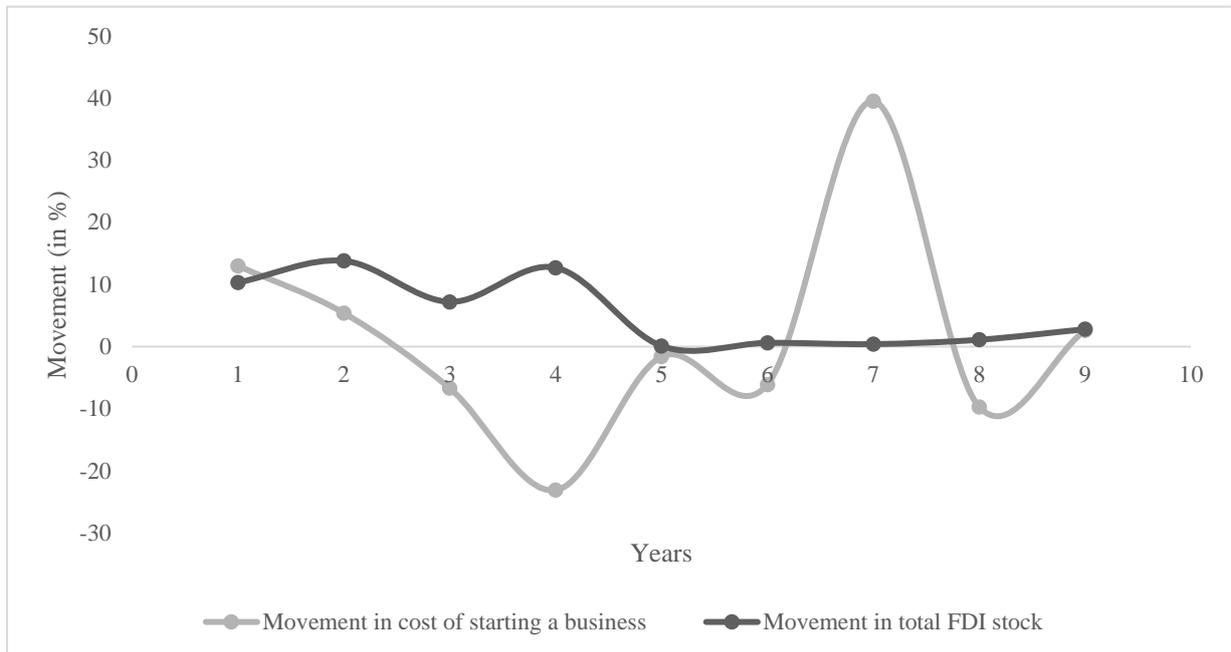

Sources: United Nations Conference on Trade and Development Data Center, 'Foreign direct investment: Inward and outward flows and stock, annual', 2019 <https://unctadstat.unctad.org/wds/TableViewer/tableView.aspx?ReportId=96740> and World Bank, *Doing Business* reports 2008-2017. Washington, D.C.: World Bank and the International Finance Corporation. <http://www.doingbusiness.org/en/reports/global-reports/doing-business-2019>. Calculations by author.

As displayed in Figure 3, the Central African Republic decreased its cost of doing business for all the years in the dataset with the exception of 2009, 2010, 2015 and 2017. Throughout the dataset, the variables did observe inverse correlations for five of the nine years; as the cost of doing business decreased, the total FDI stock increased: 2009, 2010, 2015 and 2017. This is in keeping with the hypothesis, and indicates a causality.

However, like South Africa, there is some spuriousness in the case of CAR; notably in 2009, 2010, 2015 and 2017 there were increases in the cost of starting a business in the country, but there was nonetheless an increase in FDI. In fact, there were increases for all the years in the dataset, regardless of movement in the cost of starting a business in the country. Further, those growths in FDI which coincided with increases in cost of starting a business were on average larger than those years which coincided with decreases in costs as the associated totals were 24.4% and 22.2% respectively.

*Democratic Republic of the Congo*

**Figure 4: Democratic Republic of the Congo**

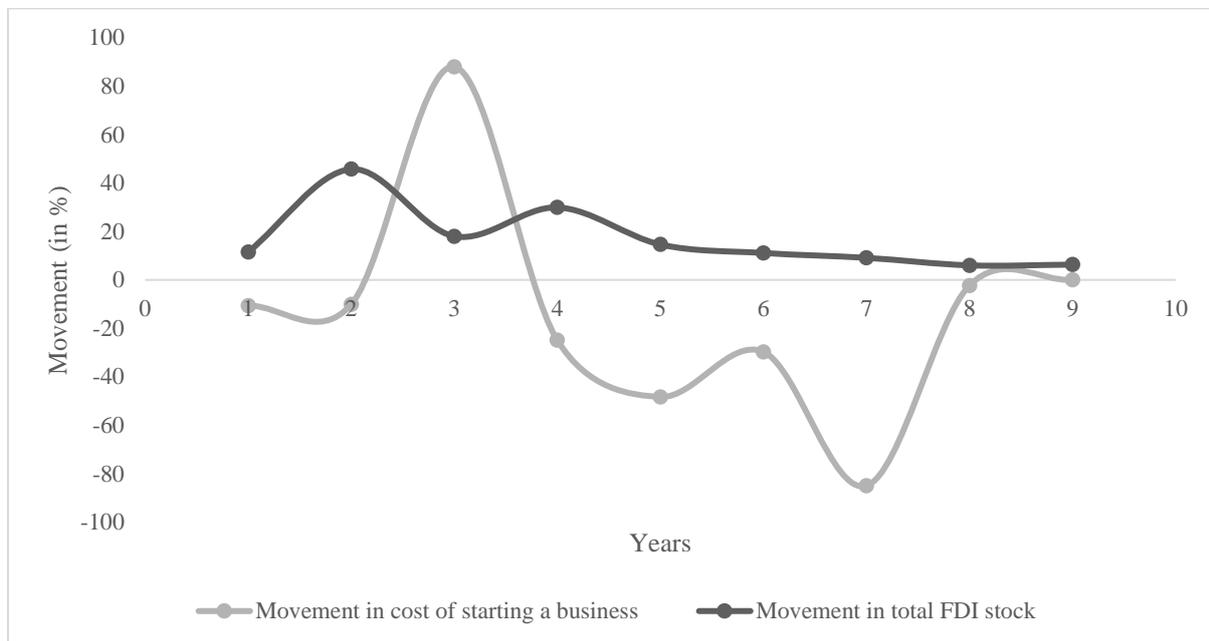

Sources: United Nations Conference on Trade and Development Data Center, 'Foreign direct investment: Inward and outward flows and stock, annual', 2019 <https://unctadstat.unctad.org/wds/TableViewer/tableView.aspx?ReportId=96740> and World Bank, *Doing Business* reports 2008-2017. Washington, D.C.: World Bank and the International Finance Corporation. <http://www.doingbusiness.org/en/reports/global-reports/doing-business-2019>. Calculations by author.

As displayed in Figure 4, the Democratic Republic of the Congo decreased its cost of doing business for all the years in the dataset with the exception of 2011. Throughout the dataset, the variables did observe inverse correlations for seven of the nine years (as much as Mauritius); as the cost of doing business decreased, the total FDI stock increased: in 2009, in 2010, in 2012, in 2013, in 2014, in 2015, and in 2016. This is in keeping with the hypothesis, and indicates a causality.

However, there were two years wherein the correlation seemed to not exist. In 2011, there was a lack of inverse correlation when the cost of starting a business increased by 88.0% but there was an increase in FDI by 18.0%. Secondly, in 2017, there neither an increase nor a decrease but there was nonetheless an increase in FDI inflow by 6.3%. This does indicate some spuriousness in the causal account; in other words, that there would be no movement in the cost of starting a business but there would be a movement in the total inward FDI stock to the DRC indicates that the *DB* index, or in the very least this variable, is not sufficient to explain the movement of FDI, at least in regards to this specific country.

**Discussion**

The overall findings from the research display a strong association between decreased costs of starting a business and FDI. But rather than it being the case that the high rankers saw the most influx of FDI, it was the case that incidences of reform (i.e., of decreases in their cost of starting a business) by them and the DRC gained the most FDI. Mauritius reduced its total cost of starting a business by 71.7% for the entire period, and saw a 167.6% in total FDI. South Africa reduced its cost of starting a business by 143.7% and saw a 79.8% increase in total FDI. And the DRC cut its cost of starting a business by 122.9% and saw a 152.21% increase in FDI.

However, there are mixed results as the movements were not universally linear. The case of CAR, which saw increased in inward FDI for all the years but only reformed for five incidences out of nine. Indeed, the country increased its cost of starting a business by 49%, and still saw an increase in FDI. However, the country saw the least growth in FDI, growing only by 13.3% and in raw figures, the country obtained the least FDI, growing only to US$651-million, from US$407-million, for the entire period. Thus, among the low rankers, in the entire timeframe the DRC proved the most active reformer and CAR overtook it as the *lower* ranking African country in the rankings.

The gains made by CAR regardless of its low pace of reform perhaps also speaks to a fundamental issue that goes back to the rationale for case selection made in selecting these countries. All four of these countries are outliers; they made up the first and final tier of African countries as ranked in the index and are perhaps guaranteed to demonstrate non-linear results. It would also be expected that they would make exponential gains as they make their reforms.

Further, the independent variable studied is one out of a total 37 variables. As the World Bank itself admits:

> The *Doing Business* methodology has limitations. Other areas important to business—such as a country's proximity to large markets, the quality of its infrastructure services (other than those related to trading across borders), the security of property from theft and looting, the transparency of government procurement, macroeconomic conditions or the underlying strength of institutions are not studied directly by *Doing Business*.[38]

The *DB's* focus on negatives (non-intrusion), and not enough focus on what is there – which may outweigh the negatives. In other words, there are internal incongruences and different directions being pulled from within the *DB* itself. For example, a state may need to increase taxes on businesses so as to expand its electricity grid. Notably, these are two variables which the *DB* encourages in different directions; reduced business taxation and access to electricity are both encouraged. This presents some level of opportunity cost for both the country as well as the foreign investors.

Further, the *DB* does not track profits being made. This essentially means that its index measures inputs and not outputs. Given that the different countries offer different primary sectors of investment, it is also possible that some countries may have costly business costs but would also result in high profit. This presents another version of the opportunity cost – the investor could pursue an investment in a country with a low cost to starting a business and potentially yield low profits; alternatively, they could pursue investment in countries with high costs of starting a business but which also prospectively offer higher profits.

There is also the factor of China, which have been both voluminous and also arguably non-traditional in its considerations and more willing to undertake risks in its investment.

> The expansion of Chinese FDI seems to be fuelled by the search for export markets (market-seeking) as FDI flows into China from developed countries have intensified over the past decades. The main reason for this type of FDI by Chinese MNCs is the intense competition faced by Chinese firms in the domestic market, due to the massive entry of FDI into the Chinese market, especially since China's entry into the World Trade Organisation, in 2001.[39]

Add Claude Sumatai and Théophile Dzaka-Kikouta:

> Under these circumstances, China does not seem to avoid the troubled areas at risk in order to increase the volume of its OFDI. In some cases, unlike its Western competitors, the Chinese state-owned MNCs have benefited from cheap capital in connection with long-term strategies. In fact, Western companies tend to consider the political instability in African countries as a constraint and this risk assessment tends to limit the opportunity to achieve high levels of FDI.[40]

**Conclusion**

This paper asked whether the World Bank's *Doing Business* index is a reliable predictor of investment influx into African states. Particularly, the paper seeks to infer whether there is a causal relationship between the various countries' scores on the cost of doing business aspect of the Doing Business index and subsequent growths or declines in FDI. In particular, the paper adopted as an independent variable the changes in scores in the World Bank's *Ease of Doing Business* index over 2009 to 2017. The overall findings from the research display a strong association between decreased costs of starting a business and FDI; the reformers (Mauritius, South Africa and the DRC) saw the most influx of FDI, as incidences of reducing costs of starting a business by them gained the most new inward FDI.

However, there are mixed results as the movements in scores and FDI across all the cases were not universally linear. The case of CAR, which saw increased in inward FDI for all the years but only reformed for five incidences out of nine, was the main exception. Indeed, the country increased its cost of starting a business by 49% in the entire period, and yet still saw an increase in FDI. However, the country saw the least amount of growth in FDI, growing only by 13.3%. In raw figures, the country obtained the least FDI, growing only to US$651-million, from US$407-million, for the entire period. Thus, among the low rankers, in the entire timeframe the DRC proved the most active reformer and CAR slid down and became the *lower* ranking African country in the rankings; though not the lowest as of 2018 due to the insertion of South Sudan. There is also the factor of China, whose investments into Africa have been both voluminous and also arguably non-traditional in their considerations and more willing to undertake risks in its investment. Future research could look at the variables and focus them specifically on Chinese FDI in these particular countries in horizontal case studies. Additional other highly-regarded indices, such as the World Economic Forum's Competitiveness Index as well as scores in the credit rankings agencies' appraisals of credit worthiness of African states, could also be put to the test in a comparative assessment of movements in scores and FDI into their own highest and lowest-ranking African countries, and ultimately distil which among them has a more accurate depiction of the determinants of new investments into African countries.

---